\documentclass[prl,showpacs,preprintnumbers,amsmath,amssymb,twocolumn,superscriptaddress]{revtex4-1}

\usepackage[dvipdfmx]{graphicx}
\usepackage{dcolumn}
\usepackage{bm}
\usepackage{color}

\usepackage{amsmath}	
\usepackage{amsfonts}
\usepackage{braket}
\usepackage{amsthm}

\newtheorem*{theorem}{Theorem}

\begin{document}

\title{
Lieb-Schultz-Mattis theorem in higher dimensions \\ 
from approximate magnetic translation symmetry
}

\author{Yasuhiro Tada}
\email[]{ytada@hiroshima-u.ac.jp}
\affiliation{
Quantum Matter Program, Graduate School of Advanced Science and Engineering, Hiroshima University,
Higashihiroshima, Hiroshima 739-8530, Japan}
\affiliation{Institute for Solid State Physics, University of Tokyo, Kashiwa 277-8581, Japan}

\begin{abstract}
We prove the Lieb-Schultz-Mattis (LSM) theorem on the energy spectrum of 
a general two or three-dimensional quantum many-body system 
with the U(1) particle number conservation and translation symmetry.
Especially, it is demonstrated that the theorem holds in a system with long-range interactions.
To this end, 
we introduce approximate magnetic translation symmetry 
under the total magnetic flux $\Phi=2\pi$
instead of the exact translation symmetry,
and explicitly construct low energy variational states. 
The energy spectrum at $\Phi=2\pi$ is shown to agree with that at $\Phi=0$ in the thermodynamic limit,
which concludes the LSM theorem.
\end{abstract}

\maketitle

{\it Introduction.}--
Understanding the low energy spectrum of a quantum many-body system is a central issue in 
condensed matter physics~\cite{Cubitt2015}.
The spectrum can be either gapless in some systems 
or it can be gapped in other systems with
spontaneously broken discrete symmetry and an intrinsic topological order~\cite{Wen,Wen2017}, 
in addition to trivial uniquely gapped systems.
In this context,
the Lieb-Schultz-Mattis (LSM) theorem is a fundamental theorem
which can put strong constraints on possible energy spectra
and provide a guiding principle for searching exotic quantum states
including topological states with long range 
entanglement
~\cite{LSM1961,AffleckLieb1986,YOA1997,Koma2000,Oshikawa2000,Hastings2004,Hastings2005,Hastings2010,NS2007,
Parameswaran2013,Watanabe2015,Lu2020,Yao2020,CGW2011,OTT2020}.
Especially, 
the original LSM theorem for one dimension holds in a system with long-range density-density interactions,
and provides a lower bound of ground state degeneracy (GSD), GSD$\geq q$, for a gapped system
with the filling per unit cell $\rho=p/q$~~\cite{LSM1961,AffleckLieb1986,YOA1997}.
The wide applicability of the theorem is fundamentally important,
since long-range interactions naturally exist in real systems~\cite{Ruelle,Lieb,Campa2009,
Schrieffer,Anderson1958,Nambu1960,
SPT_Coulomb2014,Coulomb_RMP2018,Hirata2021,DWSM2018,
FChern_LR2013,SPT_LR2013,CSL_LR2018,Dutta2015,Rydberg2020}
and they can have significant impacts
on energy spectra.
For example in three dimensions, the Coulomb interaction gaps out the collective charge excitations
in metals and plays a crucial role in the Anderson-Higgs mechanism in superconductors
~\cite{Schrieffer,Anderson1958,Nambu1960}.
Exotic quantum phases can be realized in various systems where long-range interactions are
essential, such as in
Coulomb interacting electrons~\cite{SPT_Coulomb2014,Coulomb_RMP2018,Hirata2021,DWSM2018}
and dipolar systems ~\cite{FChern_LR2013,SPT_LR2013,CSL_LR2018,Dutta2015,Rydberg2020}.
Besides, GSD is closely related to the nature of ground states for both broken discrete symmetry 
~\cite{OYA1997,Furuya2019} and a topological order~\cite{OshikawaSenthil2006,LevinWen2006,KitaevPreskill2006},
which might be affected by long-range interactions.

Unfortunately, however,
the original proof cannot be applied to a higher dimensional system with an isotropic system size, 
and higher dimensional extensions were made possible more than thirty years after the original work
~\cite{Oshikawa2000,Hastings2004,Hastings2005,Hastings2010,NS2007}.
Based on local twist of a short-range Hamiltonian
~\cite{Hastings2004,Hastings2005,Hastings2010,NS2007},
it was shown that GSD$\geq q$ for a gapped system under an assumption on matrix elements of
local operators. 
This may be generalized to some rapidly decaying long-range interacting systems,
but exact conditions are not yet known.
On the other hand, the higher dimensional LSM theorem was proved also in a different approach
under an hypothesis that an excitation gap does not close when a $2\pi$-flux quanta
piercing a hole of the torus system
is adiabatically inserted
~\cite{Oshikawa2000}. 
Although this approach is formally applicable to a system with long-range interactions,
the adiabatic hypothesis is a subtle issue especially in such a system and its validity is still under debate
~\cite{Oshikawa2000,Hastings2004,Hastings2005,Hastings2010,Misguish2002,Watanabe2018}.
Therefore, it is still not clear whether or not the LSM thereom holds in a higher dimensional system
with long-range interactions.

In this study, we discuss the LSM theorem in higher dimensions, 
especially focusing on long-range interacting systems.
With use of approximate magnetic translation
instead of the conventional one,
we can prove the theorem 
and extend its applicability to a wider class of systems.
Technically, our proof may be regarded as a simple generalization of the original one-dimensional LSM argument
and therefore long-range interactions can be treated in a straightfoward way,
which is an advantage of our approach.
To be concrete, we consider a simple model of spinless particles (either fermions or bosons) on a two-dimensional 
square lattice of a linear size $L_x\simeq L_y\simeq L=\sqrt{L_xL_y}$
with the periodic boundary condition.
Our proof is applicable also to a three dimensional system with a size $L_z\simeq L$.
The Hamiltonian is given by
\begin{align}
H(\phi)&=H_t(\phi)+H_V \nonumber \\
&=-\sum_{\langle i,j\rangle}t_{ij}(\phi)c^{\dagger}_ic_j+\frac{1}{2}\sum_{i,j}V_{ij} 
\tilde{n}_i\tilde{n}_j
\label{eq:H}
\end{align}
where $j=(x_j,y_j)$ is a site position and $\langle i,j\rangle$ represents a nearest neighbor pair of sites.
The hopping integral includes the vector potential $t_{jk}(\phi)=te^{iA_{jk}}$ with $t\in{\mathbb R}$
corresponding to a uniform magnetic flux per plaquette $\phi=\sum_{\langle i,j\rangle\in {\rm plaquette}}
A_{ij}$. 
The second term $H_V$ describes the density-density interaction with $\tilde{n}_j=c^{\dagger}_jc_j-\rho$
at the filling $\rho=p/q$
and the potential $V_{ij}=V_{|i-j|}$ can 
include long-range interactions in addition to short-range interactions.
The Hamiltonian posesses translation symmetry when $A_{ij}=0$.
We consider a class of general interactions with stability of the Hamiltonian and extensiveness of energy eigenvalues,
including stable tempered interactions and Coulomb interaction
~\cite{Ruelle,Lieb}.
Then, we prove the following statement.

\begin{theorem}
Consider the Hamiltonian $H(\phi=0)$.
When the filling per unit cell is $\rho=p/q$ with coprime $p,q\in{\mathbb N}$,
either there exist gapless excitations or the ground states are at least $q$-fold degenerate in the thermodynamic limit.
\label{thm:LSM}
\end{theorem}

The proof consists of two steps.
(i) We firstly construct approximate magnetic translation operators ${\mathcal T}_{x,y}$ in presence of $\phi_{L}
=2\pi/L_xL_y=2\pi/L^2$
and show that the low energy states of $H(\phi_{L})$ are nearly $q$-fold degenerate in a finite size system
as a consequence of a non-trivial commutation relation of ${\mathcal T}_{x},{\mathcal T}_{y}$
corresponding to a projective representation of ${\mathbb Z}\times{\mathbb Z}$.
(ii) Next, we demonstrate that the energy difference 
$\delta E_n(\Phi_{0})=[E_n(\Phi_{0})-E_n(0)]$ vanishes
in the thermodynamic limit, where $E_n(\Phi_0)$ is the $n$-th eigenvalue of $H(\phi_L)$ 
with the total magnetic flux, $\Phi_0=L_xL_y\times \phi_L=2\pi$.
By combining these two results, we can complete the proof of the main theorem~\cite{higherD}.
The proof can be generalized to a wide class of models with hopping beyond the nearest neighbors,
lattices other than the square or cubic lattice, spins and orbitals, and some other long-range interactions.
In the following, we discuss the two steps for the Hamiltonian Eq.~\eqref{eq:H} and generalizations
will be presented elsewhere.

\begin{figure}[tbh]
\includegraphics[width=2.5cm]{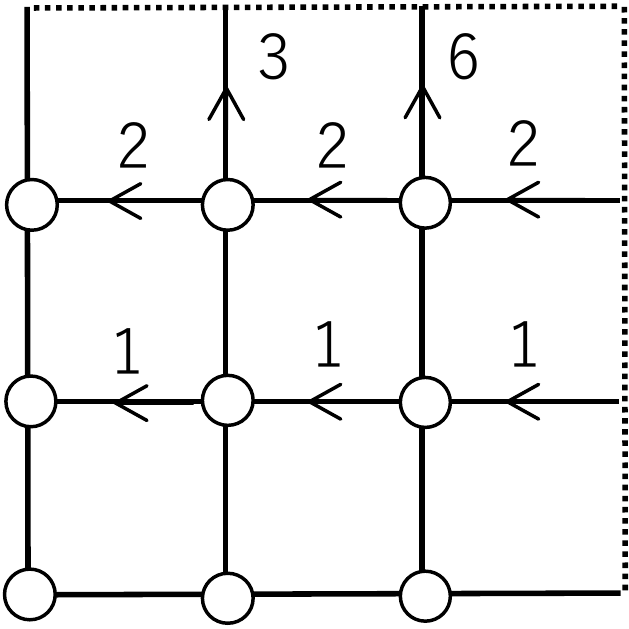}
\caption{The string gauge for a $L_x=L_y=3$ system. 
Each number on the bonds corresponds to $A_{ij}$ in unit of $\phi_{L=3}=2\pi/9$
and is given in mod 9.
}
\label{fig:string}
\end{figure}

{\it Step (i) approximate magnetic translation and low energy states.}---
Firstly, we give an explicit construction of the approximate 
magnetic translation operators for the Hamiltonian Eq.~\eqref{eq:H}
and also of low energy variational states under the small magnetic field $\phi_L$.
We consider the string gauge with the period $L_x, L_y$ 
which realizes the smallest flux per plaquette $\phi=\phi_{L}=2\pi/L^2$
and the total flux in the system $\Phi=\Phi_0=2\pi$ under the periodic boundary condition
~\cite{Hatsugai1999,Kudo2017,Tada2020}. 
In this study, 
the gauge configuration is fixed as in Fig.~\ref{fig:string} and straightforwardly generalized for arbitrary $L_x, L_y$
~\cite{Landau}.

\begin{figure}[tbh]
\includegraphics[width=8.5cm]{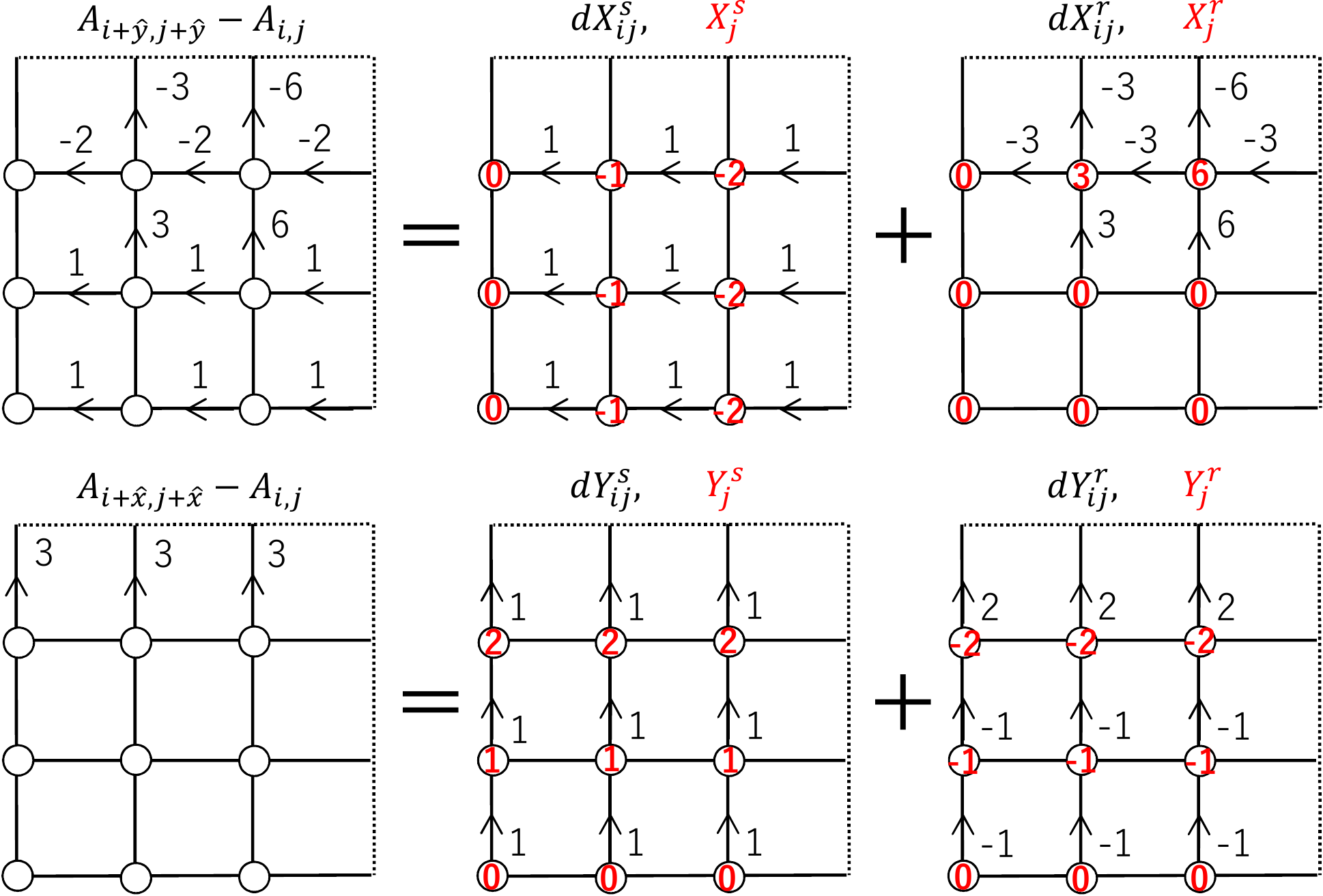}
\caption{The gauge transformation 
$A_{i+\hat{y},j+\hat{y}}-A_{i,j}=dX^s_{ij}+dX^r_{ij}$ and
$A_{i+\hat{x},j+\hat{x}}-A_{i,j}=dY^s_{ij}+dY^r_{ij}$ for $L_x=L_y=3$.
The red numbers inside the circles represent $X^{s,r}_j$ and $Y^{s,r}_{j}$.
All the numbers are defined in unit of $\phi_{L=3}=2\pi/9$ and are in mod 9.
}
\label{fig:dA}
\end{figure}

One can define an approximate magnetic translation operator in the string gauge 
by introducing appropriate scalar functions $X_j, Y_j$,
\begin{align}
{\mathcal T}_x &=T_xU_y= T_x\exp\left(i\sum_j Y_j\tilde{n}_j\right),\\
{\mathcal T}_y &=T_yU_x= T_y\exp\left(i\sum_j X_j\tilde{n}_j\right),
\end{align}
where $T_{x,y}$ are the conventional translation operators without a magnetic field.
We can determine the functions $X_j, Y_j$ by trying to require  translational symmetry of the Hamiltonian
as follows.
The hopping Hamiltonian is transformed as
\begin{align}
{\mathcal T}_{\mu}c^{\dagger}_je^{i{A}_{jk}}c_k{\mathcal T}_{\mu}^{-1}
&=c^{\dagger}_{j+\hat{\mu}}e^{iZ_j^{{\mu}}}e^{i{A}_{jk}}e^{-iZ_k^{{\mu}}}c_{k+\hat{\mu}} \nonumber\\
&\equiv c^{\dagger}_{j+\hat{\mu}}e^{i{A}_{j+\hat{\mu},k+\hat{\mu}}}c_{k+\hat{\mu}}
\label{eq:y_trans}
\end{align}
in $\mu$-direction, where $Z^x_j=Y_j, Z^y_j=X_j$.
In the second equality, we have required the magnetic translation symmetry.
This leads to the condition ${A}_{i+\hat{\mu},j+\hat{\mu}}={A}_{ij}+dZ^{{\mu}}_{ij}$ with 
$dZ^{\mu}_{ij}=Z_i^{\mu}-Z_j^{\mu}$.
This is basically a gauge transformation ${A}_{ij}\to {A}'_{ij}={A}_{i+\hat{\mu},j+\hat{\mu}}$ 
by the unknown scalar function $Z^{{\mu}}_j$.
Unfortunately, however, there is no solution for $Z_j^{\mu}$ that satisfies the simple periodic boundary condition, 
$Z_j^{x,y}=Z_{j+L_{\mu}\hat{\mu}}^{x,y}$.
We have to introduce a singular gauge transformation to satisfy
Eq.~\eqref{eq:y_trans} and correspondingly decompose $Z_j^{\mu}$ into a singular term and regular term
$Z_j^{\mu}=Z_j^{s\mu }+Z_j^{\mu r}$. 
An example of $X_j$ and $Y_j$ for $L_x=L_y=3$ is shown in Fig.~\ref{fig:dA}, 
and they are obtained in a similar way for other general
system sizes.
A singular gauge transformation is often treated with an introduction of a branch cut and
it can be explicitly implemented in our system, but we will take a different approach in this study.

Here, instead of the full magnetic translation symmetry, 
we consider only the regular parts $X^{r}_j, Y^r_j$ which approximately
realize the magnetic translation, and neglect the singular parts $X^s_j,Y^s_j$.
For simplicity, the same notation ${\mathcal T}_{x,y}$ is used for the approximated magnetic translation operator.
We stress that the regular parts alone satisfy a desired commutation relation of ${\mathcal T}_{x,y}$,
even when we ignore the singular parts correspond to a uniform singular vector potential 
$A^s_{j+\hat{\mu},j}=\phi_{\mu}$
with $\phi_x=-\phi_L,\phi_y=\phi_L$ which does not contribute to the out-of-plane flux.
Indeed, one can easilly derive the commutation relation of 
the approximate magnetic translation operator ${\mathcal T}_{x,y}$,
\begin{align}
{\mathcal T}_y^{-1}{\mathcal T}_x^{-1}{\mathcal T}_y{\mathcal T}_x =e^{i\phi_LN},
\label{eq:com}
\end{align}
where $N=\sum_jn_j=\rho L_xL_y$ at the filling $\rho$.
Therefore these operators give a projective representation of ${\mathbb Z}\times{\mathbb Z}$,
which is a key in our discussion.

Now we consider the ground state of the Hamiltonian Eq.~\eqref{eq:H} and low energy variational states.
In constructing the variational states, we use the following relations which are derived straightforwardly,
\begin{align}
{\mathcal T}_xH(\phi_L;0,0){\mathcal T}_x^{-1}=H(\phi_L;0,-\phi_y), 
\label{eq:Txcom} \\
{\mathcal T}_yH(\phi_L;0,0){\mathcal T}_y^{-1}=H(\phi_L;-\phi_x,0), 
\label{eq:Tycom}
\end{align}
where $H(\phi_L;\phi_x,\phi_y)$ is the Hamiltonian with the magnetic field $\phi_L$ along $z$-direction and
the constant vector potential $A^s_{j+\hat{\mu},j}=\phi_{\mu}$ along $\mu$-direction
with $\phi_x=-\phi_L,\phi_y=\phi_L$~\cite{twist}.
These equations mean that ${\mathcal T}_{x,y}$ describe magnetic translation symmetry
up to the small quantity $\phi_{\mu}=O(L^{-2})$, and $H$ and ${\mathcal T}_{\mu}H{\mathcal T}_{\mu}^{-1}$
are unitary equivalent with the same spectra.
In the following, we regard ${\mathcal T}_y$ as a twist operator and ${\mathcal T}_x$ as a
near symmetry operator.
Given the ground state which satisfies $H(\phi_L;0,0)\ket{\Psi_0}=E_0(\Phi_0)\ket{\Psi_0}$
for the total flux $\Phi_0=2\pi$, 
the variational states are defined by
$\ket{\Psi_{0k}}=({\mathcal T}_y)^k\ket{\Psi_0}$ with $k\in{\mathbb Z}$. 
Then, it follows from Eq.~\eqref{eq:Tycom} that 
$E_{01}(\Phi_0)
 =\langle \Psi_{01}|H(\phi_L;0,0)|\Psi_{01}\rangle 
 = \langle \Psi_{0}|H(\phi_L;\phi_x,0)|\Psi_{0}\rangle$ is evaluated as
\begin{align}
E_{01}
 = E_0 + \phi_x h_1 +\phi_x^2 h_2 + \cdots,
\label{eq:Evar}
\end{align}
where we have Taylor expanded $H(\phi_L;\phi_x,0)$ with respect to $\phi_x=O(L^{-2})$
and $h_l=\langle \Psi_{0}|\partial_{\phi_x}^lH(\phi_L;0,0)|\Psi_0\rangle/l!$. 
Clearly, 
the second correction term behaves as $\phi_x^2h_2=O(L^{-4})\times O(L^2)=O(L^{-2})$ in two dimensions.
The first correction term $\phi_xh_1$ is odd in $\phi_x$ and its sign can be flipped by considering
another variational state
${\mathcal T}_y^{-1}\ket{\Psi_0}$ in addition to ${\mathcal T}_y\ket{\Psi_0}$.
The absolute value of $\phi_xh_1$ must be smaller than that of $\phi_x^2h_2$ so that
the variational energies of $H(\phi_L;0,0)$ for the two states ${\mathcal T}_y^{\pm1}\ket{\Psi_0}$
are greater than or equal to $E_0$,
which is a variant of Bloch's theorem for the persistent current
~\cite{Bohm1949,Tada2016}.
The higher order corrections are even smaller, and
we end up with $E_{01}=E_0+O(L^{-2})$.
One also obtains $E_{0k}=E_0+O(L^{-1})$ in three dimensions.

Next, we discuss approximate orthogonality of these states based on Eq.~\eqref{eq:Txcom} which
is now regarded as a near symmetry of $H(\phi_L;0,0)$.
We first consider a case where the ground state is uniquely gapped and
later move on to a multiply degenerate case.
Following the previous study~\cite{Oshikawa2000}, we introduce a unitary evolution operator
${\mathcal F}_y$
which adiabatically inserts a flux $\Phi_y=\sum_{y_j} A^s_{j+\hat{y},j}=
L_y\phi_y$ through the non-contractible hole of the torus in $y$-direction~\cite{F}.
Since $\ket{\Psi_n(\Phi_y)}={\mathcal F}_y(\Phi_y)\ket{\Psi_n(0)}$~\cite{Kato,Kato1950},
Eq.~\eqref{eq:Txcom} leads to $H(0)\cdot {\mathcal T}_x{\mathcal F}_y\ket{\Psi_0(0)}
=E_0(\Phi_y)\cdot {\mathcal T}_x{\mathcal F}_y\ket{\Psi_0(0)}$,
where $E_0(\Phi_y)$ is the ground state energy with the flux, 
$H(\Phi_y)\ket{\Psi_0(\Phi_y)}=E_0(\Phi_y)\ket{\Psi_0(\Phi_y)}$. 
When the spectrum of $H(0)$ has a gap $\Delta(0)=O(L^0)=O(1)$ above the unique ground state,
the gap does not close for a flux $\Phi'_y\in[0,\Phi_y]$
essentially because the inserted flux $\Phi'_y=O(L^{-1})$ is vanishingly small~\cite{gap},
which implies that $E_0(0\leq \Phi'_y\leq\Phi_y)$ stays at the lowest energy.
Because the spectra of $H(0)$ and $H(\Phi_y)$ are unitary equivalent, 
this means $E_0(0)=E_0(\Phi_y)$ and hence
$\ket{\Psi_0(0)}$
is an eigenstate of the combined unitary operator ${\mathcal T}_x{\mathcal F}_y$. 
Therefore, with use of the commutation relation Eq.~\eqref{eq:com},
$\langle \Psi_0(0)|\Psi_{01}(0)\rangle=e^{-i\phi_L N}
\bra{\Psi_0(0)}({\mathcal F}_y^{-1}{\mathcal T}_x^{-1}){\mathcal T}_y({\mathcal T}_x{\mathcal F}_y)
\ket{\Psi_0(0)}+O(L^{-1})$,
we obtain in two dimensions 
\begin{align}
\langle \Psi_0|\Psi_{01}\rangle =e^{i2\pi\rho}\langle \Psi_0|\Psi_{01}\rangle +O(L^{-1}).
\label{eq:ortho}
\end{align}
To be consistent with the preassumed unique gapped ground state, 
$\rho$ must be an integer.
The contraposition corresponds to a part of the LSM theorem.
In three dimensions, the corresponding factor is $e^{i2\pi \rho L_z}$,
which also requires an integer $\rho$ for suitably chosen $L_z$
similarly to the previous study~\cite{Oshikawa2000}.

The above discussions can be extended to a gapped system with general degeneracy $D$,
from which we can conclude $D\geq q$ for $\rho=p/q$.
A fractionally filled system is either gapless or gapped with $D> 1$ as shown above, and here 
we consider the latter case 
with a gap $\Delta=O(1)$ 
from the $D$-dimensional
ground state sector to excited states for $H(\phi_L;0,0)$.
The ground state sector consists of the states $\{\ket{\Psi_n}\}_{n=0}^{D-1}$ whose 
energies agree in the thermodynamic limit and we neglect possbile vanishingly small energy differences for brevity.
Then we construct variational states $\ket{\Psi_{nk}}=({\mathcal T}_y)^k\ket{\Psi_{n}}$
for $k=1,\cdots,K$
and evaluate their energy expectation values $E_{nk}$.
We can just repeat the same argument as above and obtain
$E_{nk}=E_0+O(L^{d-4})$ in $d$-dimensions.
To discuss their (near) orthogonality, we introduce a vector
$I=(I_0,\cdots,I_{D-1})^T$ with
$I_n=\langle \Psi_{n}|\Psi_{nk}\rangle =\bra{\Psi_n}{\mathcal T}_y^k\ket{\Psi_n}$.
Then, one obtains $I=e^{i2\pi k\rho}I$ in two dimensions similarly to Eq.~\eqref{eq:ortho}
~\cite{generalGSD} 
and it sugggests
$1\leq \exists k_0\leq K$ s.t. $k_0\rho\in{\mathbb Z}$ when $K=D$
since the number of linearly independent variational states must be smaller than or equal to $D$. 
This implies $D\geq q$.


{\it Step (ii) stability of many-body eigenvalues to magnetic fields}.---
Here, we discuss stability of eigenvalues $E_n(\Phi=0)$ of $H(\phi=0)$ to a small magnetic field
in $z$-direction,
and show that $\delta E_n(\Phi_0)=[E_n(\Phi_{0})- E_n(0)]\to 0$ as $L\to\infty$.
One of the  difficulties in discussing such stability is that the uniform magnetic field $\phi_{L}$ is not 
a small perturbation in the usual sense, and $|e^{iA_{jk}}-1|$ is not vanishing for a large number of bonds,
which prevents us from Taylor expanding the Hamiltonian only up to a small finite order in $\phi_{L}$.
It is non-trivial whether or not  $\phi_L=2\pi/L^2$ can be simply regarded as 
the $\phi\to0$ limit,
since the corresponding total flux $\Phi_0=2\pi$ is $O(1)$, which could potentially lead to
$\delta E_n(\Phi_0)=O(1)$.

On the other hand, 
one may naively expect the stability of the many-body eigenvalues, $\delta E_n(\Phi_{0})\to 0$,
as has been assumed in numerical calculations~\cite{Assaad2002}.
To explicitly demonstrate it, 
we use the stability of single-particle eigenvalues $\varepsilon_n(\phi=0)$ to a magnetic field,
which was mathematically proved in the literature~\cite{Berkolaiko2013,Berkolaiko2014,Verdiere2013}.
To use this result, we have to appropriately modify our Hamiltonian by
introducing an on-site potential term $H_U=\sum_iU_i n_i$ which can lift the degeneracy of the
single-particle eigenvalues.
Here, we choose $U_j$ to be a fixed random potential in $[-u,u]$ for a given system size
so that the degeneracy of $\varepsilon_n(\phi=0)$ 
due to spatial (rotation, inversion, and translation) symmetries is lifted.
Besides, the corresponding single-particle eigenfunctions will be non-zero anywhere in the system, 
because of the random potential which suppresses accidental zeros.
Then, one has $\delta \varepsilon_n(\phi_L)=[\varepsilon_n(\phi_L)-\varepsilon_n(0)]\sim \phi_L^2=O(L^{-4})$ 
possibly with a $u$-dependent coefficient
~\cite{Berkolaiko2013,Berkolaiko2014,Verdiere2013}.

This immediately leads to eigenvalue stability
of the non-interacting Hamiltonian $H_{tU}(\phi_{L},u)=H_{t}(\phi_{L})+H_U(u)$, namely,
$\delta E_n(\Phi_0,u,V=0)\sim \phi_{L}^2N=O(L^{d-4})$ in $d$-dimensions.
We keep $u>0$ to show $\delta E_n(\Phi_{0},u)\to 0$ in the thermodynamic limit,
and then turn off the random potential, $u\to 0$~\cite{uL},
which eventually implies $\delta E_n\to 0$ in absence of the artificial potential $U_j$.
We can also see that corresponding changes in eigenvectors of $H_{tU}(\phi_L,u)$ are 
vanishingly small; 
a direct calculation gives $\|\ket{\delta\Psi_n(\phi_L,u)}\|^2=\| \ket{\Psi_n(\phi_L,u)}-\ket{\Psi_n(0,u)}\|^2
=O(\phi_L^2 N)=O(L^{d-4})$.
Therefore the eigenvalue stability implies that
the resolvent $R_{tU}(\phi_L,u;E)=[H_{tU}(\phi_L,u)-E]^{-1}$ approaches $R_{tU}(0,0;E)$
in the above mentioned limit.

Now we consider eigenvalue stability of the interacting Hamiltonian $H(\phi_L,u)=H_{tU}(\phi_L,u)+H_V$.
We can see that the eigenvalues and eigenvectors of $H(\phi_L,u)$ approach those at $\phi=0$ in a similar manner.
This follows from the resolvent equation
\begin{align}
&[H_{tU}(\phi_L,u)+H_V-E]^{-1}\nonumber\\
&\qquad =[H_{tU}-E]^{-1}[1+H_V[H_{tU}-E]^{-1}]^{-1},
\label{eq:resolvent}
\end{align}
where
$[H_{tU}(\phi_L,u)-E]^{-1}\to [H_{tU}(0,u)-E]^{-1}$ as already discussed.
Therefore, we conclude $[H_{tU}(\phi_L,u)+H_V-E]^{-1}\to [H_{tU}(0,u)+H_V-E]^{-1}$, 
which means stability of the eingevalues and eingevectors of
$H(\phi_L,u)=H_{tU}(\phi_L,u)+H_V$ to the small magnetic field $\phi_{L}$
at $u\neq 0$.
Finally, we take the limit $u\to 0$ and conclude that the eigenvalues of the clean many-body Hamiltonian
for the sufficiently large system approach $E_n(\Phi=0)$.
Since the eigenvectors of $H(\phi_L)$ also converge to those of $H(0)$,
the (near) orthogonality Eq.~\eqref{eq:ortho} is kept down to $\phi=0$.
This completes our proof of the LSM theorem.

In summary, with use of the approximate magnetic translation symmetry,
we have extended the LSM theorem to higher dimensional long-range interacting systems
and derived the lower bound, ${\rm GSD}\geq q$,
for gapped ground state degeneracy at a fractional filling $\rho=p/q$.

We are grateful to Y. Yao, M. Oshikawa, A. Ueda, T. Koma, M. G. Yamada, M. Sato, S. C. Furuya, K. Shiozaki,
and S. Kamimoto for valuable discussions.
This work was supported by JSPS KAKENHI Grant
No. JP17K14333.

\bibliography{ref_}

\begin{thebibliography}{60}
\bibitem [{hig()}]{higherD}%
  \BibitemOpen
  \href@noop {} {}\bibinfo {note} {In this sense, the flux $\phi_L$ is regarded
  as a perturbation to control the system in two or three dimensions. It would
  be an interesting future study to consider a different perturbation
  controlling a system in dimensions higher than three.}\BibitemShut {Stop}%
\bibitem [{twi()}]{twist}%
  \BibitemOpen
  \href@noop {} {}\bibinfo {note} {The twists in the right hand side arise from
  the approxiation of neglecting the singular terms. Indeed, with the
  application of ${\mathcal T}_x$ to the Hamiltonian $H(\phi_L;0,0)$, the
  hopping term becomes
  $c_{j+\hat{x}}^{\dagger}e^{iA_{jk}+idY^r_{jk}}c_{k+\hat{x}} =
  c_{j+\hat{x}}^{\dagger}e^{iA_{j+\hat{x},k+\hat{x}}-idY^s_{jk}}c_{k+\hat{x}}$,
  where $dY^s_{jk}=A^s_{jk}=\phi_y$ itself is well-defined although $Y^s_j$ is
  not. A similar argument holds for ${\mathcal T}_y$.}\BibitemShut {Stop}%
\bibitem [{F()}]{F}%
  \BibitemOpen
  \href@noop {} {}\bibinfo {note} {{Note that ${\mathcal F}_y=1+O(L^{-1})$
  essentially because $\Phi_y=O(L^{-1})$. Indeed, the ${\mathcal F}_y(\Phi_y)$
  can be written as ${\mathcal
  F}_y(\Phi_y)=\hat{T}\exp(-i\int_0^{\Phi_y}Q(\Phi_y')d\Phi_y')$, ${\mathcal
  F}_y(\Phi_y)=1-i\int_0^{\Phi_y}Q(\Phi_y')d\Phi_y'+\cdots$, where $Q(\Phi_y)$
  is given by $Q=i\sum_n\partial_{\Phi_y}P_nP_n$ with the projection operator
  $P_n(\Phi_y)=\ket{\Psi_n}\bra{\Psi_n}$ for the $n$-the eigenstate of
  $H(\Phi_y)=H(\phi;0,\Phi_y/L_y)$~\cite{Kato,Kato1950}. In this expression,
  the Hermitian operator $Q$ is an $O(1)$ operator whose operator norm is
  simply $\|Q\|=O(1)$ and nearly independent of the system size, which implies
  ${\mathcal F}_y=1+O(\Phi_y)=1+O(L^{-1})$. This is in contrast to the
  $2\pi$-flux insertion discussed previously~\cite{Oshikawa2000} for which the
  corresponding unitary operator might be non-trivial.}}\BibitemShut {Stop}%
\bibitem [{gen()}]{generalGSD}%
  \BibitemOpen
  \href@noop {} {}\bibinfo {note} {More precisely, after the adiabatic time
  evolution where the gap remains non-zero similarly to the case with $D=1$, a
  state $\ket{\Psi_n}$ may change to another state
  $\sum_{n=0}^{D-1}P_{mn}\ket{\Psi_n}$ in the ground state sector with a
  unitary matrix $P$. So, we change the basis states such that the unitary
  matrix $P$ is diagonalized, and rewrite them as
  $\{\ket{\Psi_n}\}_{n=0}^{D-1}$ using the same symbols for simplicity. In this
  basis, one obtains $I=e^{i2\pi k\rho}I$ in two dimensions.}\BibitemShut
  {Stop}%
\bibitem [{uL()}]{uL}%
  \BibitemOpen
  \href@noop {} {}\bibinfo {note} {More precisely, $\forall u>0, \forall
  \epsilon>0, \exists L_0>0 \textrm{ s.t. } |\delta E_n(\Phi_0,u)|<\epsilon
  \textrm{ for } L>L_0$. Note that the order of the double limits
  $\lim_{u\to0}\lim_{L\to\infty}\delta E_n$ is not
  interchangeable.}\BibitemShut {Stop}%
\end{thebibliography}

\end{document}